\documentclass[12pt]{article}
\usepackage{amssymb}
\usepackage{epsfig}
\usepackage{graphicx}
\usepackage{amsmath}
\usepackage{verbatim}
\usepackage{titlesec}

\csname @addtoreset\endcsname{equation}{section}
\begin{document}
\begin{titlepage}
\title{\bf\Large Effective Field Theory Analysis on  $\mu$ Problem in Low-Scale Gauge Mediation \vspace{18pt}}
\author{\normalsize Sibo~Zheng  \vspace{15pt}\\
{\it\small  Department of Physics, Chongqing University, Chongqing 401331, P.R. China}\\}

\date{}
\maketitle \voffset -.3in \vskip 1.cm \centerline{\bf Abstract}
\vskip .3cm Supersymmetric models based on the scenario of gague
mediation often suffer from the well-known $\mu$ problem. In this paper,
we reconsider this problem in low-scale gauge mediation in terms of
effective field theory analysis. In this paradigm, all high energy input soft mass can be expressed via loop expansions.  If the corrections coming from messenger thresholds are small, as we assume in this letter, then all RG  evaluations
can be taken as linearly approximation for low-scale supersymmetric
breaking. Due to these observations, the parameter space can be
systematically classified and studied after constraints coming from
electro-weak symmetry breaking are imposed. We find that some old
proposals in the literature are reproduced, and two new classes are uncovered.
We refer to a microscopic model,
where the specific relations among coefficients in one of the new classes are well motivated.
Also, we discuss some primary phenomenologies.

\vskip 5.cm
\noindent May 2011
\thispagestyle{empty}

\end{titlepage}
\newpage
\section{Introduction}
\subsection{Motivations}
Among scenarios that solve the stringent constraints on experiments of flavor violations in supersymmetric (SUSY) models,
gauge mediation \cite{GM1, GM2, GM3, GM4, GM5} is an appealing candidate.
In particular,  SUSY models with low-scale SUSY-breaking can be directly examined at the LHC.
However, gauge mediation suffers from a well-known fine tuning problem related to electro-weak symmetry breaking (EWSB), i.e,
the $\mu$ and B$\mu$ problem \cite{mu}. On general grounds there is
a small hierarchy between these two input scales at mediation scale, $B_{\mu}\sim~ (16\pi^2)\mu^{2}$.
There have been a lot of efforts on solving this problem in the literature \cite{old}.
However, except all important subtleties are taken into account,
it is not concrete to argue whether a SUSY model suffers such a fine tuning problem.

What is the most important essentials related to $\mu$ problem are all the input soft masses that are determined
both by the SUSY-breaking and the messenger sectors as well as the renormalization group (RG) evaluations
for these soft masses.
The specific relations in the mediation scale can be substantially modified by the RG corrections.
A typical example is SUSY models of gaugino mediation \cite{gaugino},
in which the high energy input parameters are three gaugino masses (together with $L^{-1}$, $\tan\beta$ and $sign{\mu}$).
It obviously does not realize EWSB and satisfy other considerations.
But the correct and large RG corrections turns it into a viable SUSY model at electro-weak scale.

\subsection{Strategy}
In this paper, we systematically analyze the $\mu$ problem from viewpoint of effective field theory,
where the SUSY-breaking sector is referred by spurion superfields.
We introduce two separate spurion superfields, which generalize the minimal setup of gauge mediation where
only one spurion field needs to be considered. These spurion superfields introduce relevant dynamical scales,
\begin{eqnarray}{\label{e1}}
X_{G}=M_{G}\left(1+\theta^{2}\Lambda_{G}\right),
~~~~~~X_{D}=M_{D}\left(1+\theta^{2}\Lambda_{D}\right)
\end{eqnarray}
$X_G$ and $X_D$ are responsible for SUSY-breaking related to generations of scalar scalars, gaugino masses as well as
$\mu$ and $B_{\mu}$ terms of Higgs superfields, respectively.
In this note, for simplicity we assume $M_{D}/M_{G}\sim 0.1-1$ so that  the corrections coming from multiplet messenger thresholds are small.

Using $X_{G}$ and $X_{D}$, we can express all the high energy input soft mass terms through loop expansions
\footnote{We take the effective number of messengers as $N_{eff,G}=1$ for the squark and gaugino masses.
This value can vary when the models derive from the minimal setup (see e.g, \cite{EOGM}).
It thus changes the relative ratio between squark and gaugino masses. }.
Note that the specific relations $\Lambda_{G}/M_{G}\lesssim 10^{-1}, \Lambda_{D}/M_{D}\lesssim 10^{-1}$ have to be imposed such
that we can express
\begin{eqnarray}{\label{e2}}
m^{2}_{\tilde{f}_{i}}&=&\sum^{3}_{i=1}\left(C_{2}(\tilde{f},i)
\frac{g^{4}_{i}}{(16\pi^{2})^{2}}\right)\Lambda^{2}_{G},\nonumber\\
M_{i}&=&\frac{g^{2}_{i}}{(16\pi^{2})}\Lambda_{G}
\end{eqnarray}
for squark and gaugino masses and
\begin{eqnarray}{\label{e3}}
m^{2}_{{H_{\mu},d}}=\sum^{2}_{i=1}\left(C_{2}(H_{\mu,d},i)
\frac{g^{4}_{i}}{(16\pi^{2})^{2}}\right)\Lambda^{2}_{G}
+\Lambda^{2}_{D}\left[\frac{C^{(1)}_{H_{\mu,d}}}{(16\pi^{2})}
+\frac{C^{(2)}_{H_{\mu,d}}}{(16\pi^{2})^{2}}+\cdots\right],
\end{eqnarray}
for Higgs masses squared $m^{2}_{H_{\mu}}$ and $m^{2}_{H_{d}}$.
The contributions in \eqref{e3} arise from  the ordinary gauge mediation and the superpotential for hidden sector $X_D$.
The $\mu$ and B$\mu$ terms, on the other hands, only receive the contributions coming from $X_D$-sector
\footnote{Note that a tree-level $\mu$ term can exist in the SUSY limit $\Lambda_{D}=0$.},
\begin{eqnarray}{\label{e4}}
B_{\mu} &=&\Lambda^{2}_{D}\left[\frac{C^{(1)}_{B\mu}}{(16\pi^{2})}
+\frac{C^{(2)}_{B\mu}}{(16\pi^{2})^{2}}+\cdots\right]\nonumber\\
\mu&=&C^{(0)}_{\mu}M_{D}+\Lambda_{D}\left[\frac{C^{(1)}_{\mu}}{(16\pi^{2})}
+\frac{C^{(2)}_{\mu}}{(16\pi^{2})^{2}}+\cdots\right],
\end{eqnarray}
Note that when a particular loop coefficient $C^{(i)}$ is determined to be non-zero,
all higher order corrections can be neglected because of the perturbative nature.
Furthermore,  a coefficient $C^{(i)}$ smaller than loop factor $1/(16\pi^{2})$ is actually an $(i+1)$-loop
effect. Following this fact we assume all $C^{(i)}$ are bounded as,
\begin{eqnarray}{\label{e5}}
\frac{1}{16\pi^{2}}< C^{(i)}\lesssim 1
\end{eqnarray}
The upper bound in \eqref{e5} is due to the perturbativity of new Yukawa couplings in superpotential for hidden sector $X_D$.
Otherwise, large Yukawa couplings will suffer the problem of Landau poles.

Having systematically analyzed the high energy input soft masses,
we can obtain their values at EW scale by taking the RG corrections into account.
The RG equations \cite{RGE} for these soft masses in minimal supersymmetric standard model (MSSM) are quite involved.
In this note we discuss gauge mediation with low-scale SUSY breaking,
\begin{eqnarray}{\label{a1}}
10^{2}TeV\lesssim\Lambda_{G,D}\lesssim 10^6TeV
\end{eqnarray}
In this region the NLSP particle are mainly prompt decays or can be long lived,
which is of highly interest at searches of SUSY signals at colliders.
Also in the region \eqref{a1} the RG evaluations for soft masses mainly receive their
corrections coming from $\ln(M/M_{Z})$ factor, which is of order unity .
This contribution can be taken as linear approximation \cite{9609434}.
Following the RG equations at one-loop,
one obtains the soft masses at EW scale via replacing \eqref{e3} and \eqref{e4} with
\begin{eqnarray}{\label{e6}}
\mu\rightarrow\mu&=&\hat{\mu},\nonumber\\
B_{\mu}\rightarrow ~B_{\mu}-\frac{\delta C_{B_{\mu}}}{(16\pi^{2})^{2}}\mu\Lambda_{G}&=&\hat{B_{\mu}},\nonumber\\
m^{2}_{H_{\mu}}\rightarrow~m^{2}_{H_{\mu}}-\frac{\delta C_{H_{\mu}}}{(16\pi^{2})^{2}}\Lambda^{2}_{G}&=&\hat{m}^{2}_{H_{\mu}},\\
m^{2}_{H_{d}}\rightarrow~m^{2}_{\tilde{H}_{d}}+\frac{\delta C_{H_{d}}}{(16\pi^{2})^{2}}\Lambda^{2}_{G}&=&\hat{m}^{2}_{H_{d}}\nonumber
\end{eqnarray}
where $\delta~ C_i$ are both positive real numbers of order one
\footnote{The $\ln(M)$ terms are all of order one in low-scale gauge mediation,
which are absorbed into the $\delta~C_{i}$.}.

Then, as a primary analysis about the physical parameter space,
 we impose the necessary condition for EWSB,
\begin{eqnarray}{\label{e7}}
\left(\hat{m}^{2}_{H_{\mu}}+\mid\hat{\mu}\mid^{2}\right)
\left(\hat{m}^{2}_{H_{d}}+\mid\hat{\mu}\mid^{2}\right)\thickapprox \hat{B}^{2}_{\mu}
\end{eqnarray}
In terms of \eqref{e6} and expressions for input soft masses through loop expansions,
we can analyze the possible parameter space composed of
$(C^{(i)}_{\mu}, C^{(i)}_{B_{\mu}}, C^{(i)}_{H_{\mu,d}}, M_{G,D}, \Lambda_{G,D})$ under constraints \eqref{e5} order by order.
An important observation which can be used to simplify the classifications is that
if $C^{(i)}_{\mu}=0$ at $i$th-loop, then we must also have $C^{(i)}_{B_{\mu}}=0$.
Reversely, the statement is not true.
This conclusion can be proved by arguments of effective field theory
(see \cite{prove} for primary analysis via SUSY algebra).

Other relevant experimental constraints on parameters in Higgs sector include negative mass squared for $H_{\mu}$ scalar,
\begin{eqnarray}{\label{a2}}
\hat{m}^{2}_{H_{\mu}}<0
\end{eqnarray}
all masses of Higgs scalars are bigger than $\mathcal{O}(115)$GeV, and
the mass of the lightest chargino has to be larger than $\mathcal{O}(100)$GeV.
We will also discuss possible implications arising from these constraints on allowed parameter space.

\subsection{Outlines}
The outline of this paper is organized as follows.
In section 2, we divide the discussions into three classes,
\begin{eqnarray}{\label{e8}}
Case~(1)&:&~~~~\hat{\mu}^{2}>>\mid\hat{m}^{2}_{H_{\mu}}\mid, ~~\hat{\mu}^{2}>>\hat{m}^{2}_{H_{d}}\nonumber\\
Case~(2)&:&~~~~\hat{\mu}^{2}> \mid\hat{m}^{2}_{H_{\mu}}\mid, ~~\hat{\mu}^{2}<<\hat{m}^{2}_{H_{d}}\nonumber\\
Case~(3)&:&~~~~\hat{\mu}^{2}>\mid\hat{m}^{2}_{H_{\mu}}\mid, ~~\hat{\mu}^{2}\sim \hat{m}^{2}_{H_{d}}\\
Case~(4)&:&~~~~\hat{\mu}^{2}<<\mid m^{2}_{H_{\mu}}\mid, ~~\hat{\mu}^{2}\sim \hat{m}^{2}_{H_{d}}\nonumber
\end{eqnarray}
First, we find that there is parameter space allowed for the case $(1)$,
which is not extensively studied before as far as we know.
There are parameter spaces allowed  for case $(2)$ to case $(4)$.
Second, the typical parameter space for case $(3)$ and case $(4)$
corresponds to the one-loop $\mu$/two-loop B$\mu$ as well as  ``lopsided gauge mediation " \cite{1103.6033} respectively,
which both have been discussed in the literature.
Finally, the second case is totally new.
Although the spectra for soft masses in Higgs sector in this case is similar to those in \cite{0809.4492} ,
it is not covered by that approach.
Here only one substantial hierarchy among soft masses in Higgs sector is needed,
in comparison with at least two proposed in \cite{0809.4492}.

In section 3, we discuss the microscopic models of hidden sector $X_D$ where the
typical values in parameter space for case $(2)$ is well motivated.
The new Yukawa couplings appearing in the hidden superpotential $W(X_{D},H_{\mu,d})$
are all around unity or smaller.
No severe fine tunings are allowed in the hidden sector.
This is also different from the model buildings for the approach in \cite{0809.4492}.

In section 4, we discuss the phenomenologies for models belonging to case $(2)$.
Following the typical parameter values, we find that Higgs scalar expect $h^{0}$ are nearly degenerate at large
$\hat{m}_{H_{d}}$ and small $\tan\beta\thickapprox 0.1$ is favored.
The direct consequence for this class of SUSY models is that the next-to-lightest supersymmetric particle (NLSP) is mostly
Bino-like and its prompt two-body decay modes are dominated by final state $\gamma$ and  $Z^0$ plus missing energy,
with branching ratio $cos^{2}\theta_{W}$ and $sin^{2}\theta_{W}$ respectively.
The decay channel to $h^0$ plus missing energy is negligible as a result of dramatically suppression.

In section 5, we summarize the main results in this note.

\section{Classifications}
Now we discuss the allowed parameter space in the four classes in \eqref{e8} in turn.
First,we introduce such three dimensional parameters for later use,
\begin{eqnarray}{\label{e9}}
x=\Lambda_{G}/M_{G}\lesssim 10^{-1},~~~~ y=\Lambda_{D}/M_{D}\lesssim 10^{-1},~~~~~z=\Lambda_{G}/\Lambda_{D}.
\end{eqnarray}
The first two are small positive, real numbers as mentioned in the introduction,
while the last parameter $z$ is also real and positive, in the range of $10^{-2}$ to $10^{2}$ in general.
When $z=1$, we reproduce the physics of gauge mediation with one spurion superfield.
Since the mass parameters in Higgs sector receive contributions both from the $X_G$ and $X_D$ sectors
either directly or via RG evaluation, it will be of use to discuss which one is the dominant contribution.

\subsection{Case $(1)$: Large $\hat{\mu}$ Term}

In this case, the most stringent constraint comes from \eqref{e7}
, from which we have when $\mu$ is generated at tree-level
\begin{eqnarray}{\label{e10}}
\left(C^{(0)}_{\mu}\right)^{2}&\thickapprox&\left[\frac{C^{(1)}_{B\mu}}{(16\pi^{2})}y^{2}
+\frac{C^{(2)}_{B\mu}}{(16\pi^{2})^{2}}y^{3}+\cdots\right]-\frac{C^{(0)}_{\mu}\delta C_{B_{\mu}}}{(16\pi^{2})^{2}}\frac{\Lambda_{G}}{M_{D}}
\end{eqnarray}
First, consider the case for $C^{(1)}_{B_{\mu}}\neq 0$. Eq.\eqref{e10} implies that,
\begin{eqnarray}{\label{e11}}
\left(C^{(0)}_{\mu}\right)^{2}\thickapprox\frac{C^{(1)}_{B_{\mu}}}{16\pi^{2}}y^{2},~~~~~~~~~~~~
C^{(1)}_{B_{\mu}}>\frac{C^{(0)}_{\mu}\delta C_{B_{\mu}}}{16\pi^{2}}\frac{M_{D}}{\Lambda_{G}}
\end{eqnarray}
Second, if $C^{(1)}_{B_{\mu}}= 0$ and $C^{(2)}_{B_{\mu}}\neq 0$, then one obtains,
\begin{eqnarray}{\label{e12}}
\left(C^{(0)}_{\mu}\right)^{2}\thickapprox\frac{C^{(1)}_{B_{\mu}}}{(16\pi^{2})^{2}}y^{3}
\end{eqnarray}
which is not allowed according to \eqref{e5}.
Note that $\delta C_{B_{\mu}}$ is a positive real coefficient of order one.
As the tree-level $\mu$ term respects the supersymmetry,
it is less of interest in comparison with the mechanism
that all mass scales are originated from softly broken SUSY,
unless a dynamical generation for tree-level $\mu$ term is realized so that it is a natural consequence.

If $\mu$ is generated at one-loop, \eqref{e10} is replaced by,
\begin{eqnarray}{\label{e13}}
\left(C^{(1)}_{\mu}\right)^{2}&\thickapprox&16\pi^{2}\left[C^{(1)}_{B_{\mu}}
+\frac{C^{(2)}_{B_{\mu}}}{(16\pi^{2})}y+\cdots\right]-\frac{C^{(1)}_{\mu}\delta C_{B_{\mu}}}{(16\pi^{2})}
\frac{\Lambda_{G}}{\Lambda_{D}}
\end{eqnarray}
A non-zero $C^{(1)}_{B_{\mu}}$ will cause the tension for its permitted value, as stated in \eqref{e5}.
What is more promising is that the $B_{\mu}$ term is generated at two-loop.
In this case, we obtain,
\begin{eqnarray}{\label{e14}}
C^{(1)}_{\mu}>z,~~~~~~~~~~~~
\left(C^{(1)}_{\mu}\right)^{2}\thickapprox yC^{(2)}_{B_{\mu}}
\end{eqnarray}
Take the constraints for case $(1)$ in \eqref{e8} into account,
we find that no matter the mainly contribution to $\hat{m}_{H_{\mu,d}}$ comes from $X_D$ or $X_G$ sector,
there is only a set of critical values,
\begin{eqnarray}{\label{e15}}
(z_{*})^{2}=\frac{1}{16\pi^2}, ~~~~~~and~~~~~~C^{(1)}_{\mu*}=1
\end{eqnarray}
if we relax the large differences in \eqref{e8} for case $(1)$ to some moderate values.

\subsection{Case $(2)$: Large $\hat{m}_{H_{d}}$}
We obtain after imposing the constraint in \eqref{e8} for this case ,
\begin{eqnarray}{\label{e18}}
\begin{array}{c}
   \mid  C_{H_{\mu,d}}^{(2)}\mid\lesssim z^{2}\\
  C^{(0)}_{\mu}>\frac{x}{(16\pi^{2})}\\
C^{(0)}_{\mu}<<\frac{x}{(16\pi^{2})}\\
\cdots\\
\end{array}
~~~~~~or~~
\begin{array}{c}
   \mid  C_{H_{\mu}}^{(2)}\mid< z^{2}\\
   C_{H_{d}}^{(2)}>> z^{2}\\
C^{(0)}_{\mu}>\frac{x}{(16\pi^{2})}\\
\left(C^{(0)}_{\mu}\right)^{2}<<\frac{ C^{(2)}_{H_{d}}}{(16\pi^{2})}y^{2}\\
\cdots\\
\end{array}
~~~~~~or~~~
\begin{array}{c}
   \mid C_{H_{\mu}}^{(2)}\mid > z^{2}\\
  \mid C_{H_{d}}^{(2)}\mid<< z^{2}\\
C^{(0)}_{\mu}<<\frac{x}{(16\pi^{2})}\\
\left(C^{(0)}_{\mu}\right)^{2}>\frac{ C^{(2)}_{H_{d}}}{(16\pi^{2})}y^{2}\\
\cdots\\
\end{array}
\end{eqnarray}
if $C^{(0)}_{\mu}\neq 0$. Here the constraints neglected denote those coming from \eqref{e7}.
We see that the first and third sets of choices is obviously not consistent.
The second one implies that $\ C^{(2)}_{H_{d}}y>>1$,
which is not permitted as a result of the fact that
$ C^{(2)}_{H_{d}}$ is of order unity and $y$ is smaller than one.

if $C^{(0)}_{\mu}=0$ and $C^{(1)}_{\mu}\neq 0$, we get
\begin{eqnarray}{\label{e19}}
\begin{array}{c}
   \mid C_{H_{d}}^{(2)}\mid\lesssim z^{2}\\
  \left(C^{(1)}_{\mu}\right)^{2}>x^{2}\\
\left(C^{(1)}_{\mu}\right)^{2}<<x^{2}\\
\cdots\\
\end{array}
~~~~or~~
\begin{array}{c}
   \mid  C_{H_{\mu}}^{(2)}\mid\lesssim z^{2}\\
   C_{H_{d}}^{(2)}>> z^{2}\\
\left(C^{(1)}_{\mu}\right)^{2}\gtrsim z^{2}\\
\left(C^{(1)}_{\mu}\right)^{2}<<C_{H_{d}}^{(2)}\\
\cdots\\
\end{array}
~~~~~~or~~~
\begin{array}{c}
   \mid  C_{H_{\mu}}^{(2)}\mid> z^{2}\\
   C_{H_{d}}^{(2)}<< z^{2}\\
\left(C^{(1)}_{\mu}\right)^{2}> C_{H_{d}}^{(2)}\\
\left(C^{(1)}_{\mu}\right)^{2}<<z^{2}\\
\cdots\\
\end{array}
\end{eqnarray}
The first and third classes are obviously not consistent,
while the second one is allowed.
We impose the constraint in \eqref{e7}, which implies that when $C^{(1)}_{B\mu}= 0$,
$\sqrt{C^{(2)}_{H_{d}}}\thickapprox z/(16\pi^{2})>> z$.
Thus this choice is not allowed.
On the other hand, if $C^{(1)}_{B\mu}\neq 0$, we obtain an additional constraint,
\begin{eqnarray}{\label{e20}}
C^{(1)}_{\mu}\sqrt{C^{(2)}_{H_{d}}}\thickapprox C^{(1)}_{B\mu}>>C^{(1)}_{\mu}z
\end{eqnarray}
which is actually consistent with the second class of choices in \eqref{e19}.
For example, we take such typical values for these parameters,
\begin{eqnarray}{\label{e21}}
C^{(1)}_{\mu}\sim C^{(1)}_{B\mu}\sim z\sim  0.1, ~~~C^{(2)}_{H_{\mu}}\sim 0.01~~~
and~~~~~C^{(2)}_{H_{d}}\sim 1
\end{eqnarray}
In the next section, we will construct a class of models to motivate choices of parameters in \eqref{e21}.

\subsection{Case $(3)$: Standard Proposal}
First, when $C^{(0)}_{\mu}\neq 0$, the only possible parameter space corresponds to
the circumstance under which the contributions to soft masses of $H_{\mu,d}$ are
dominated by the $X_G$ hidden sector. Otherwise, we will always get extremely small $C^{(0)}_{\mu}$, i.e,
$C^{(0)}_{\mu}<<1/(16\pi^{2})$, which indeed is high-order effects according to our understandings.
Follow this observation, we obtain the constraints,
\begin{eqnarray}{\label{e22}}
\begin{array}{c}
   \mid C_{H_{\mu}}^{(2)}\mid\lesssim z^{2},\\
   \mid C_{H_{d}}^{(2)}\mid\lesssim z^{2},\\
  C^{(0)}_{\mu}\thickapprox\left(\frac{x}{16\pi^{2}}\right),\\
\left(C^{(0)}_{\mu}\right)^{2}\thickapprox\frac{C^{(1)}_{B\mu}}{(16\pi^{2})}y^{2}\\
\end{array}
\end{eqnarray}

Second, if $\mu$ term is generated at one-loop, there is only one possibility needed to be considered,
\begin{eqnarray}{\label{e23}}
\begin{array}{c}
   \mid C_{H_{\mu}}^{(2)}\mid< z^{2},\\
   \mid C_{H_{d}}^{(2)}\mid< z^{2},\\
  C^{(1)}_{\mu}\thickapprox z^{2},\\
\left(C^{(1)}_{\mu}\right)^{2}\thickapprox 16\pi^{2} C^{(1)}_{B\mu}
~~~or~~~\left(C^{(1)}_{\mu}\right)^{2}\thickapprox yC^{(2)}_{B\mu}, ~~~(C^{(1)}_{B\mu}=0)\\
\end{array}
\end{eqnarray}
for either one-loop or two-loop  $B_{\mu}$. All other choices are directly excluded.
These two choices have been well known.
As shown in \eqref{e23}, for the one-loop generation of $B_{\mu}$ ,
there will be no parameter space allowed except $C^{(2)}_{B_{\mu}}$ close to its lower bounded value,
which actually a two-loop effect.
In this sense,
the second class in \eqref{e23} is often referred as the standard proposal to solve the problem.

\subsection{Case $(4)$: Small $\hat{\mu}$ and Large $\hat{B_{\mu}}$}
Similarly to the previous discussions, the fist class of choices for $C^{(0)}_{\mu}\neq 0$ is given by,
\begin{eqnarray}{\label{e24}}
 C_{\mu}^{(0)}<< \frac{1}{16\pi^{2}}\frac{\Lambda_{G}}{M_{D}},~~~~~~~
C_{\mu}^{(0)}<< \frac{1}{16\pi^{2}}y
\end{eqnarray}
which contradicts with the statement in \eqref{e5}.
Second, if $C^{(0)}_{\mu}=0$ and $C^{(1)}_{\mu}\neq 0$,
we obtain,
\begin{eqnarray}{\label{e25}}
 C_{\mu}^{(1)}<< z,~~~~~~~~~~~~~~~~~
C_{\mu}^{(1)}<< C^{(2)}_{H_{d}}
\end{eqnarray}
These choices are argued to be consistent in a class of hidden model \cite{1103.6033}.

In summary, we reproduce two old proposals in the literature as well as discover two new approaches.
In particular , the proposal discussed in \cite{0809.4492} is a specific choice of the case $(2)$. This statement will be more apparent in the next section, in which the model buildings and the parameter space is obviously different from the discussions shown in  \cite{0809.4492}.

\section{A Hidden Model}
 In this section we consider the hidden superpotential for SUSY-breaking sector $X_D$ associated with
 the Higgs sector. We refer to the hidden superpotential as
 \footnote{This superpotential is the one in \cite{1103.6033}
 that has been used to analyze the models of large $B_{\mu}$ and small $\mu$.
 Here we use it for different purpose and as an illustration for constructing mass spectrum in \eqref{e21}.
 It is also interesting to discuss other possibilities such as models of MSSM singlets. }
\begin{eqnarray}{\label{e30}}
W_{hid}=\lambda_{\mu}H_{\mu}DS+\lambda_{d}H_{d}\bar{D}\bar{S}+X_{D}D\bar{D}+\frac{1}{2}X_{D}
\left(a_{S}S^{2}+a_{\bar{S}}\bar{S}^{2}+a_{S\bar{S}}S\bar{S}\right)
\end{eqnarray}
in terms of which the typical parameter space in \eqref{e21} for the case $(2)$ can be realized.
In \eqref{e30}, $D$ and $\bar{D}$ are bi-fundamental chiral superfileds under the representation of the EW groups,
$S$ and $\bar{S}$ are MSSM singlets, and
 $X_{D}$ and $X_S$ refer to the relevant hidden SUSY-breaking sector.

Integrate out the massive component fields in $S,\bar{S},D$ and $\bar{D}$,
we obtain the effective Kahler potential \cite{Kahler},
\begin{eqnarray}{\label{e31}}
K_{eff}=-\frac{1}{32\pi^{2}}Tr~\left[\mathcal{M}\mathcal{M}^{\dag}
\ln\left(\frac{\mathcal{M}\mathcal{M}^{\dag}}{\Lambda^{2}}\right)\right]
\end{eqnarray}
where
\begin{eqnarray}{\label{e32}}
\mathcal{M}\mathcal{M}^{\dag}=\left(
\begin{array}{cc}
\mid X_{D}\mid^{2}+\mid\lambda_{d}\mid^{2}\mid~H_{d}\mid^{2} & \lambda_{\mu}H_{\mu}X_{D}^{*}+\lambda_{d}H_{d}^{*}X_{S} \\
(\lambda_{\mu}H_{\mu}X_{D}^{*}+\lambda_{d}H_{d}^{*}X_{S})^{*} & \mid X_{S}\mid^{2}+\mid\lambda_{\mu}\mid^{2}\mid~H_{\mu}\mid^{2} \\
\end{array}
\right)
\end{eqnarray}
The soft masses can be obtained via evaluating the eigenvalues of matrix \eqref{e32}, which are given by
\footnote{For the explicit expressions for functions $P(v,w),Q(v,w)$ and $R(v,w)$ we refer the readers to \cite{1103.6033}.},
\begin{eqnarray}{\label{e33}}
m^{2}_{H_{\mu}}&=&\frac{\lambda_{\mu}^{2}}{16\pi^{2}}\Lambda_{D}^{2}
\left[c_{\theta}^{2}P(v,w)+s_{\theta}^{2}P(\lambda~v,w)\right],\nonumber\\
m^{2}_{H_{d}}&=&\frac{\lambda_{d}^{2}}{16\pi^{2}}\Lambda_{D}^{2}
\left[s_{\theta}^{2}P(v,w)+c_{\theta}^{2}P(\lambda~v,w)\right],\\
\mu&=&\frac{\lambda_{\mu}\lambda_{d}}{16\pi^{2}}\Lambda_{D}s_{\theta}c_{\theta}\left[Q(\lambda~v,w)-Q(v,w)\right],\nonumber\\
B_{\mu}&=&\frac{\lambda_{\mu}\lambda_{d}}{16\pi^{2}}\Lambda^{2}_{D}s_{\theta}c_{\theta}\left[R(\lambda~v,w)-R(v,w)\right]\nonumber
\end{eqnarray}
where the dimensionless coefficients are defined as,
\begin{eqnarray}{\label{e34}}
\lambda=\frac{a_{\bar{S}}-a_{S}\tan^{2}\theta}{a_{S}-a_{\bar{S}}\tan^{2}\theta},
~~~~\tan2\theta=\frac{2a_{S\bar{S}}}{a_{\bar{S}}-a_{S}},~~~~~
v=M_{S}/M_{D},~~~~w=\Lambda_{S}/\Lambda_{D},
\end{eqnarray}
Compare \eqref{e3} and \eqref{e4} with \eqref{e33}, we obtain the explicit expressions,
\begin{eqnarray}{\label{e35}}
C^{(1)}_{\mu}&=&\lambda_{\mu}\lambda_{d}s_{\theta}c_{\theta}\left[Q(\lambda~v,w)-Q(v,w)\right],\nonumber\\
C^{(1)}_{B_{\mu}}&=&\lambda_{\mu}\lambda_{d}s_{\theta}c_{\theta}\left[Q(\lambda~v,w)-Q(v,w)\right],\nonumber\\
C^{(2)}_{H_{\mu}}&=&(16\pi^{2})\lambda_{\mu}^{2}\left[c_{\theta}^{2}P(v,w)+s_{\theta}^{2}P(\lambda~v,w)\right],\\
C^{(2)}_{H_{d}}&=&(16\pi^{2})\lambda_{d}^{2}\left[s_{\theta}^{2}P(v,w)+c_{\theta}^{2}P(\lambda~v,w)\right]\nonumber
\end{eqnarray}

\begin{figure}[!h]
\centering
\begin{minipage}[b]{0.6\textwidth}
\centering
\includegraphics[width=4in]{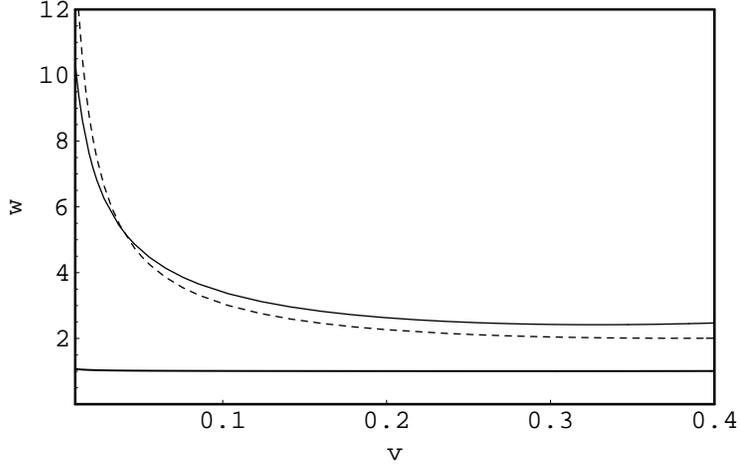}
\end{minipage}%
\caption{ The values of $w$s vary as $v$s in the region where $\sin\theta\sim 0.02$, with
$\lambda_{\mu}=4.8$, $\lambda_{d}=0.4$ and $\lambda=0.1$.
The dashing, solid and the bottom lines represent the curves for $\mu$, $B_{\mu}$ and $H_{\mu}$ respectively.
The favored region is near $v\sim 0.2-0.3$.
In this region $C^{(2)}_{H_{d}}\sim 0.4-0.6$.}
\end{figure}

Now we discuss the parameter space that is composed of $\lambda_{\mu,d}, \lambda, v$, $\theta$
and $ w$ ( with an overall scale $\Lambda_{D}$ ),
by impose the concrete choices in \eqref{e21} on \eqref{e35}.
In particular, the perturbativity of Yukawa couplings in \eqref{e30} and the absence
of obvious fine tunings between them require,
\begin{eqnarray}{\label{e36}}
 0.1\lesssim\lambda_{\mu,d}\lesssim 5,~~~~~~~
0.1\lesssim\lambda\lesssim 10,~~~~~~~~~0.1\lesssim a_{S,\bar{S}}\lesssim 5,
\end{eqnarray}
Also the value of $v$ is restricted to be around $0.1-10$ according to the assumption we follow.

First, we take the curves for $C^{(1)}_{\mu}$ and $C^{(2)}_{B_{\mu}}$ for consideration.
In the region of small $\sin\theta$ value where substantial simplifications happen,
the large ratio of $\lambda_{\mu}/\lambda_{d}$ is favored. If $\lambda$ is set to be around $0.8$,
these two curves overlap from $v\sim0.1$ to $v\sim 0.3$,
during which $C^{(2)}_{H_{\mu}}$ is close to its typical value, as shown in fig. $(1)$.

When we move to the region of parameter space where $\cos\theta$ closes to 1,
similar results can be obtained, with $a_{S}\sim 0.1 a_{\bar{S}}$ and $a_{S\bar{S}}\sim 0.1 a_{\bar{S}}$ as shown in fig $(2)$.
In this region, it is more easier to obtain large $m_{H_{d}}$ term.
The typical value for $C^{(1)}_{H_{d}}$ is around $1.0-3.0$.

The main goal of this section is to construct a concrete model in which the typical parameter space is allowed.
We want to mention that the parameter space we discover is part of the total physical one.

\begin{figure}[!h]
\centering
\begin{minipage}[b]{0.6\textwidth}
\centering
\includegraphics[width=4in]{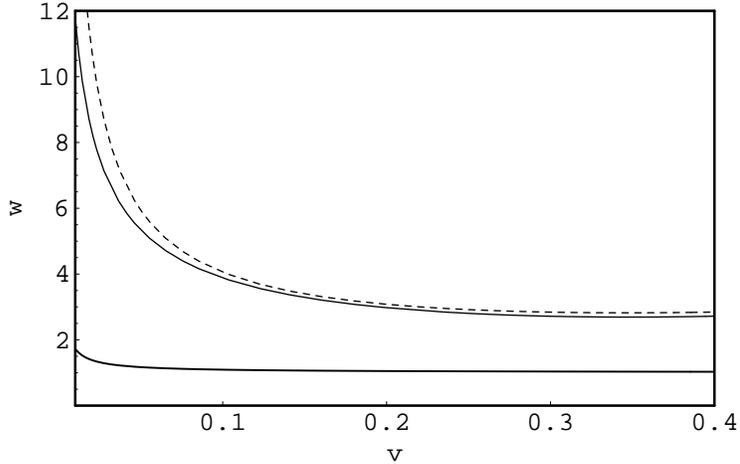}
\end{minipage}%
\caption{ The values of $w$s vary as $v$s in the region where $\cos\theta\sim 0.02$, with
$\lambda_{\mu}=4$, $\lambda_{d}=0.4$ and $\lambda=0.1$.
The dashing, solid and the bottom lines represent the curves for $\mu$, $B_{\mu}$ and $H_{\mu}$ respectively.
The favored region is near $v\sim 0.3-0.4$.}
\end{figure}

\section{Phenomenology}
Now we discuss the phenomenological implications for the mass spectra represented by the second class of models in \eqref{e8}.
In the typical parameter space,  we find,
\begin{eqnarray}{\label{e37}}
\mid\hat{m}^{2}_{H_{\mu}}\mid<\hat{\mu}^{2}\sim M^{2}_{r}\sim \hat{m}^{2}_{\tilde{f}}
\sim 10^{-1}\hat{B}_{\mu}\sim 10^{-2}\hat{m}^{2}_{H_{d}}
\end{eqnarray}

\subsection{Higgs Mass Spectrum}
The tree-level Higgs mass spectra are given by,
\begin{eqnarray}{\label{e38}}
\hat{m}^{2}_{A^{0}}=\frac{2\hat{B}_{\mu}}{\sin2\beta}=2\mid \mu\mid^{2}+\hat{m}^{2}_{H_{\mu}}+\hat{m}^{2}_{H_{d}}\thickapprox \hat{m}^{2}_{H_{d}}
\end{eqnarray}
The last expression is obtained in terms of specific relations in \eqref{e37}.
Eq.\eqref{e38} also implies that $\tan\beta\thickapprox 0.1$.
The masses of the remaining neutral Higgs scalars are,
\begin{eqnarray}{\label{e39}}
\hat{m}^{2}_{h^{0},H^{0}}=\frac{1}{2}\left(\hat{m}^{2}_{A^{0}}+m_{Z}^{2}\mp \sqrt{(\hat{m}^{2}_{A^{0}}-m_{Z}^{2})^{2}+4m_{Z}^{2}\hat{m}^{2}_{A^{0}}\sin^{2}2\beta}\right)
\end{eqnarray}
from which we obtain up to leading order of $O(m^{2}_{Z}/\hat{m}^{2}_{H_{d}})$,
\begin{eqnarray}{\label{e40}}
\hat{m}_{h^{0}}\thickapprox\frac{1}{\sqrt{2}}m_{Z}+\mathcal{O}\left(\frac{m^{2}_{Z}}{\hat{m}^{2}_{H_{d}}}\right),
~~~~~~~~~\hat{m}_{H^{0}}\thickapprox \hat{m}_{H_{d}}+\mathcal{O}\left(\frac{m^{2}_{Z}}{\hat{m}^{2}_{H_{d}}}\right)
\end{eqnarray}
The masses for charged Higgs scalars are,
\begin{eqnarray}{\label{e41}}
\hat{m}_{H^{\pm}}\thickapprox \hat{m}_{H_{d}}+\mathcal{O}\left(\frac{m^{2}_{Z}}{\hat{m}^{2}_{H_{d}}}\right)
\end{eqnarray}
These mass spectra suggest that $\Lambda_{G}\gtrsim 10^{2}$TeV.

\subsection{Bino-like NLSP}
The NLSP is either the neutrilino $\tilde{\chi}^{0}_{1}$ or chargino $\tilde{\chi}^{+}$ for the class of models in \eqref{e38}.
Using the fact that
$m_{Z}<\mid \hat{\mu}\pm M_{1}\mid$ and $m_{Z}<\mid \hat{\mu}\pm M_{2}\mid$ for $\Lambda_{G}\gtrsim 10^{2}$TeV,
we have
\begin{eqnarray}{\label{e42}}
\hat{m}_{\tilde{\chi}^{0}_{1}}&\thickapprox& M_{1}+\mathcal{O}\left(\frac{m^{2}_{W}}{\hat{\mu}^{2}}\right),\nonumber\\
\hat{m}^{2}_{\tilde{\chi}_{+}}&\thickapprox& M^{2}_{2}+\frac{1}{2}m^{2}_{W}+\mathcal{O}\left(\frac{m^{2}_{W}}{\hat{\mu}^{2}}\right)
\end{eqnarray}
So neutrilino is the NLSP. Under the same limit,
the matrix $N$ \cite{neutrilino} that is used to dialogize the neurtilino mass matrix is given by  ,
\begin{eqnarray}{\label{e43}}
N_{11}&\thickapprox&1,\nonumber\\
N_{12}&\thickapprox&0,\nonumber\\
N_{13}&\thickapprox& s_{W}s_{\beta}\frac{m_{Z}(\hat{\mu}+M_{1}\cot\beta)}{\mid\hat{\mu}\mid^{2}-M_{1}^{2}}\\
N_{14}&\thickapprox& -s_{W}c_{\beta}\frac{m_{Z}(\hat{\mu}+M_{1}\tan\beta)}{\mid\hat{\mu}\mid^{2}-M_{1}^{2}}\nonumber
\end{eqnarray}
which leas to the conclusion that the NLSP neutrilino is mostly Bino-like.
It promptly decays into final states of $\gamma$, $Z$ and $h$ plus missing energy when $\Lambda_G\lesssim 10^3$TeV,
which can be directly produced at Tevatron and LHC
( see \cite{0911.4130} for recent discussions on neutrilino NLSP at colliders).
The decay widths for these final states are given by,
\begin{eqnarray}{\label{e44}}
\Gamma(\tilde{\chi}^{0}_{1}\rightarrow \gamma+\tilde{G})&\thickapprox&c^{2}_{W}\mathcal{A},\nonumber\\
\Gamma(\tilde{\chi}^{0}_{1}\rightarrow Z+\tilde{G})&\thickapprox& \left[s_{W}^{2}+\frac{1}{8}\sin^{2}2\beta\left(\frac{m_{Z}}{\hat{\mu}}\right)^{2}
\left(1-\frac{m^{2}_{Z}}{\hat{m}^{2}_{\tilde{\chi}^{0}_{1}}}\right)^{4}\right]\mathcal{A}\thickapprox s^{2}_{W}\mathcal{A},\nonumber\\
\Gamma(\tilde{\chi}^{0}_{1}\rightarrow h+\tilde{G})&\thickapprox& 0
\end{eqnarray}
with
\begin{eqnarray}{\label{e45}}
\mathcal{A}=\frac{\hat{m}^{5}_{\tilde{\chi}_{0}}}{16\pi M^{2}_{G}\Lambda^{2}_{G}}
=\left(\frac{\hat{m}_{\tilde{\chi}_{0}}}{100GeV}\right)^{5}\left(\frac{100TeV}{\sqrt{M_{G}\Lambda_{G}}}\right)^{4}\frac{1}{0.1 mm}
\end{eqnarray}
To derive the final results in \eqref{e44} we use the definite value $\tan\beta\thickapprox 0.1$, as imposed by \eqref{e38}.
This small value dramatically suppresses the decay width of channel $\tilde{\chi}^{0}_{1}\rightarrow h+\tilde{G}$.
The branching ratios for the channels involved are found to be,
\begin{eqnarray}{\label{e46}}
Br(\tilde{\chi}^{0}_{1}&\rightarrow
& \gamma+\tilde{G})\thickapprox 74 \%\nonumber\\
Br(\tilde{\chi}^{0}_{1}&\rightarrow
& Z+\tilde{G})\thickapprox 26 \% 
\end{eqnarray}

The spectra \eqref{e37} lead to a large $m_{A}$ (relative to $m_{Z}$) and small $\tan\beta$.
No matter the value of $\tan\beta$, 
large $m_{A}$ is sufficient to reduce the Higgs search of MSSM to the SM Higgs search. 
In this sense, in the model we study here phenomenologies related 
to the rates of Higgs prodcutions and decays
are similar with those under the decoupling limit \cite{decouple}. 
However, as reviewed in \cite{0208209}, 
it is still possible to discover or exclude heavy $ m_{H^{\pm}}$ up to 1 TeV via $\tau$ $\mu$ decays at LHC with 300 fb$^{-1}$.

To descriminate between our model and the others described by the decoupling limit,
one should turn to analyze the neutralino and chargino sectors.
Due to the smaller $\mu$ term as well as smaller gaugino masses than $m_{A}$ as shown in \eqref{e37},
lighter neutrilinos and charginos are allowed,
in comparison with what one expects in the ordinary models that solve the $\mu$ problem, 
with $\mu\sim m_{H}\sim m_{A}$.
Typically, if the masses of neutrilinos and charginos are of order $\mathcal{O}(100)$ GeV, 
the decays with leptonic final states such as $\tilde{C}^{\pm}_{2}\rightarrow \tilde{N}_{1}l^{\pm}\nu$ 
and $\tilde{N}_{2}\rightarrow \tilde{N}_{1}l^{+}l^{-}$ are very interesting \cite{Martin}. 
Because these decays can be prompt and searched directly at the LHC . However, in ordinary models with $\mu\sim m_{H}\sim m_{A}$ ,  one expects the masses of neutralinos and charginos are heavier than $\mathcal{O}(200)$ GeV if we take the decoupling limit. 
A double increase in $m_{\tilde{\chi}_{0}}$  will results in  the decay widths $\Gamma(\tilde{\chi}^{0}_{1}\rightarrow \gamma/Z+\tilde{G})$ dramatically enhanced,  as shown from \eqref{e45},
which is beyond the reach of LHC.

\section{Conclusions}
In this paper $\mu$ problem in gauge mediation is studied in terms of effective field theory analysis.
As the subtleties that determine the SUSY soft mass parameters at EW scale are  the high energy input boundary values
and their RG evaluations, we use the loop expansions to capture the mainly property of the former contribution,
and restrict our discussion to negligible multiple messenger threshold corrections as well as low-scale SUSY-breaking to simplify the later contribution.
Following these facts, we classify the problem into four classes,
and explore the parameter space allowed by imposing primary constraints coming from EWSB.
We reproduce two old proposals in the literature as well as discover two new classes.

As illustration for model buildings,
we refer to a hidden theory proposed in \cite{1103.6033} for different purpose.
We find in some regions where no fine tunings happen
these typical values of parameters in one of the two new classes can be realized.
Also, we discuss the phenomenological implications predicted by this new class of models.
The Higgs scalars expect $h^0$ are heavy and of order $m_{H_{d}}$.
The NLSP is Bino-like neutrilino, whose two-body decays are prompt and mainly composed of $\gamma$ and $Z$ plus missing
energy. The channel for neutrilno decaying into $h^0$ is kinetically permitted,
however, dramatically suppressed.\\

~~~~~~~~~~~~~~~~~~~~~~~~~~~~~~~~~~~~~~~~
$\bf{Acknowledgement}$\\
This work is supported in part by the
Fundamental Research Funds for the Central Universities with project
number CDJRC10300002.

\end{document}